\documentstyle[twoside,fleqn,espcrc2,epsf]{article}

\newcommand{\AmS}{{\protect\the\textfont2
  A\kern-.1667em\lower.5ex\hbox{M}\kern-.125emS}}

\title{ Nonperturbative real time propagation at finite temperature}

\author{E. Mendel \thanks{Presented at Lattice97 by E. Mendel.} and M. Nest\\
 { FB Physik,
 Carl von Ossietzky Universit\"at Oldenburg,
         26111 Oldenburg, Germany}}      
\begin{document}

\begin{abstract}
  A new formalism will be presented in order to study real
time evolution of quantum systems at finite temperature. 
Probability distributions for time-correlated observables
will be studied non-perturbatively and fully quantized. 
This works for various  systems, including ones with
tunneling phenomena.  We have obtained good results with
some computational methods which can be used on
models with several degrees of freedom.
Thus it looks feasible to study vacuum tunneling in real
time for relevant field theories at finite T.
\end{abstract}

\maketitle

\section{Introduction}

          We have been working on a
non-perturbative approach to combine real time and finite 
Temperature for the full quantized theory. This is crucial
for the study of systems with tunneling at intermediate temperatures.  

  We choose to work directly with projectors onto initial and
final states, given that the system at finite $T$ is
described by the density matrix $\rho$. This method gives a direct
probability interpretation for going from states where some
degrees of freedom have some values at time $0$, to some other
values at time $t$ (in the presence of the thermal bath produced
by all the degrees of freedom). This procedure is physically
appealing as it multiplies probabilities to be in quantum states
(given by $\rho$) by the probabilities to go from these states 
(projected onto some states for the subsystem) to some states
at later time.  This contrasts with the usual    $< \exp(- \beta H) \ [Q(t)-Q(0)]^2 >$
to estimate time correlations \cite{Smit}, where the Operator to measure
is positive by construction but one has the problem 
that it is hard to interpret as a transition probability.
 
 The  probabilities to find the system in some configuration 
can be written in a simple way 
in terms of the path integral formulation in "complex" time.
  We have been able to solve several systems
like the h.osc., the double well, $(1-\cos x)$ potential or the 
interesting case of a ring of $N$ coupled oscillators, with the 
aim of the field theory case with vacuum tunneling.

\section{Finite time formalism}

   Usually, at finite $T=1/\beta$, one just takes the trace over Euclidean time
$\beta$ to calculate expectation values. For time correlations we will 
need  a path in the complex time plane of the shape:
$\ (0,0) \rightarrow (t,-\epsilon) \rightarrow (0,-2\epsilon) \rightarrow (0,-\beta) $.\\
  We will combine finite $T$ and real $t$ non-perturbatively, by inserting 
suitable operators at the points $(0,0)$, $(t,-\epsilon)$ and $(0,-2\epsilon)$. 
With a similar path Semenoff,
Niemi, Weiss and others\cite{Seme} deduced  Feynman perturbation rules.

  To fix ideas, let us ask for the probability that if the first particle
(of a system of $N$ interacting  quantum particles) is located close to $x_1$
at $t=0$, it will be close to $y_1$ at time $t$,  the system being at 
$T = 1/\beta$. 
  We will see that this is given by:
$$
P(x_1,y_1)\!=\! \! {\int}\! dx_{2}..  dx'_{2}.. dy_{2}.. 
\ G_{t}(x_{1},x_{2},..;y_{1},y_{2},..)  
$$
\begin{equation}
 \cdot \ G^{-1}_{t}(y_{1},y_{2},..;x_{1},x'_{2},..)
 \ G_{\beta}(x_{1},x'_{2},..;x_{1},x_{2},..)     
\end{equation}
  In order to deduce  this equation, we start from the knowledge that for
$t=0$ the system is in a {\sl mixed state}, described by the density matrix:
\begin{equation}
            \rho = \exp(-\beta H) = \sum_n  \exp(-\beta E_n) |n> <n| 
\end{equation}
with the additional condition that the first degree of freedom be 
at $x_1$, which is obtained by applying the projector 
$ P_{x_1}:= |x_1><x_1| *  1\!{\rm I} _{ 2,..N} \ \ $
 on both sides
of $\rho$.  So, for $t=0$ we describe the system by 
$P_{x_1}\rho P_{x_1}$.
This operator evolves as usual, 
$ U^+_t ( P_{x_1} \rho P_{x_1} ) U_t$ , with $ U_t = \exp(-i H t)$ .

We measure again at a time $t$, with a position $y_1$ for the first 
particle.  The probability is then:  
\begin{equation}
   P_t^{\beta} (x_1,y_1) = \frac{1}{\rm Norm.} Tr[ P_{y_1} U^+_t ( P_{x_1} \rho P_{x_1} ) U_t ]    
\end{equation}
where we have discarded one of the  $P_{y_1}$  due to the cyclicity of 
the trace.
  Note that we have to take a small window around $x_1$  as otherwise we 
would measure the position with infinite precision implying total delocalization 
afterwards (we can see this effect numerically, in that the $P(x_1,y_1)$ goes to a 
constant in $y_1$ for very small windows). This can be done easily by defining the
projector on a window around the coordinate $x_1$: 
$\int_{x_1-a}^{x_1+a} dx_1 |x_1><x_1| \ $,
which still has $P_{x_1}^2 = P_{x_1}$. 
The Normalization can be taken so that the integration over $y_1$ in Eq.(3) is one,
corresponding to the assumption that the particle {\em is} in the window $[x_1-a, x_1+a]$,
or with $Z$, in which case we include the a priori probability to be in that window.
In both cases the integrated $P$ is $t$-independent.  

  Notice that  classically $\hat{P}_{x_1}$ can be commuted with $\hat{\rho}$ 
so that Eq.(3) would resemble a time
correlation among projectors at finite $\beta$, but in the quantum case this 
ordering gives a probability interpretation, while neglecting one of 
the $P_{x_1}$ one is mixing amplitudes
for different states in the thermal mixture, which one should not do...

   We can recast  Eq.(3)  to show that it corresponds to the sum of
probabilities to be in states $ |n> $ with position $ x_1$ for one of the degrees 
of freedom by the probability to have gone to $ y_1 $ in  time $ t $ :
\begin{eqnarray}
P(x_1,y_1) = \int_{i=2,..,n} \! \! dy_i \sum_n \exp(-\beta E_n)   \nonumber \\
\ \ \cdot \ \ |\int_{i=2,..,n} \! \! dx_i  G_t(y_1,y_i;x_i,x_i) \psi_n (x_1,x_i)|^2
\end{eqnarray}                                            
($G_t(x;y):=<y|U_t|x>$ is the Green function and $\psi_n (x)$ a wave function) as we wanted to prove. 

Technically it is convenient to rewrite Eq.(3) as
$$
P(x_1,y_1)=\int_{i=2,..,n} \!  \! \!  \! \! \! dx_i dx'_i dy_i <x_1,x_i|\rho |x_1,x'_i>
$$             
\begin{equation}
\ \ \ \ \cdot <x_1,x'_i|U^+_t|y_1,y_i><y_1,y_i|U_t |x_1,x_i>,
\end{equation}
a product of three Green functions as announced in Eq.(1).
The point is that each of these Green functions can be calculated 
with the method of multiplying the transfer matrix for infinitesimal "$dt$" by
itself n times to obtain times $ t=dt*(2^n)$. This method works even for 
real $t$
as we showed in our work with Dullweber and Hilf \cite{Mend}. Alternatively, one 
can implement a Fast Fourier Transform method \cite{Onof} to calculate the
short time propagator and  iterate for longer times. Both of these 
methods get improved nicely when introducing a small imaginary time
$i \epsilon$ in the time propagation, as then the propagators are  
well behaved at large distances (suppressing the fast oscillations).
  With this $ i \epsilon$ we have even been able to use normal 
Monte Carlo methods in real time, as now there is some weighting
in the sampling, obtaining satisfactory results.  
  The generalization to a
field theory seems immediate, but technically it will be hard to 
implement even on coarse lattices due to the number of degrees of freedom.
\section{Solved models}
  We have tested these methods for the systems:

a) for a free particle, the probability to find the particle at some
distance $|y_1-x_1|$ away has a near to gaussian shape, getting broader with
time. As we higher the  $T$, the dispersion grows faster.

b) for the harmonic oscillator,as shown in fig.1, from the original 
window $-.75 \pm .5$ for $t=0$, it 
oscillates with semiclassical frequency maintaining
its coherence (as known for h.osc.). With higher $T$ the packet
is broader but stays coherent.

\begin{figure}[hb]
\vspace{-6mm}
\hspace{-6mm}
\epsfxsize=80mm
\epsffile{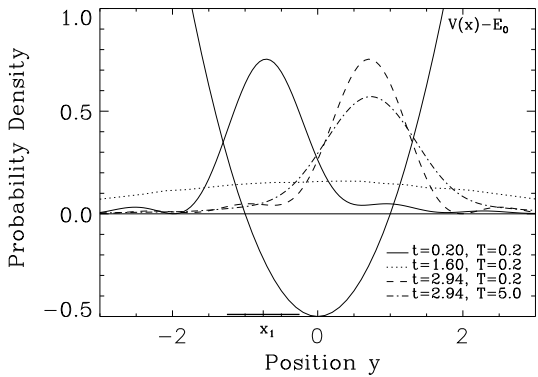}
\vspace{-10mm}
\caption{Probability  to be at $y$ for an h.osc. for times $t \approx$ 
$0$, $\pi/2$ and $\pi$, at low and high $T$.} 
\end{figure}

\begin{figure}[t]
\hspace{-6mm}
\epsfxsize=80mm
\epsffile{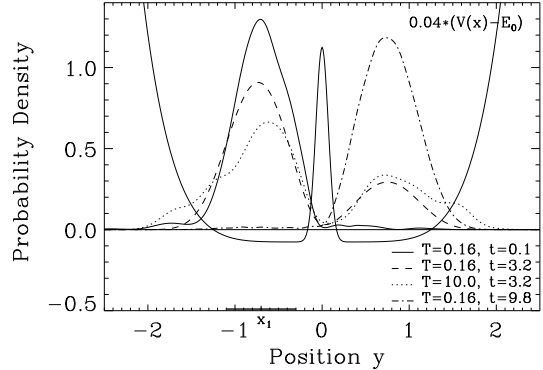}
\vspace{-10mm}
\caption{Probability  to be at $y$ for a double well for $t$ $\approx$ 
$0$, $.3$ and $1.$ of tunneling time, for two $T$.} 
\vspace{-4mm}
\end{figure}
c) the double-well potential shown in fig.2,
 where the original packet placed at one
side disperses through tunneling to the other minima. At higher
$T$ the probability to be in the other minima increases, as expected.
 We are also studying the $(1-cos(x))^n$ potentials with multiple minima
to see the dispersion with $t$.
 This case is relevant, as similar tunneling via instantons in the
field theory model is thought to produce the anomaly responsible for
baryon  asymmetry in the universe.

d) The most interesting case that we have been able to solve, is the
N particle chain of coupled oscillators in a circle, with transversal
oscillation. This is the first genuine finite temperature system that
we treated in the sense that the Temperature can be dynamically generated
by the system in a self consistent way (in contrast with an external 
given heat bath).

We have been able to solve this model "almost" analytically by going to 
normal modes, $x_i \rightarrow X_s$. In these new coordinates the Hamiltonian
is separable and the N-particle Green function is just
$$ 
<X_1,.,X_N;t|X'_{1},.,X'_{N};0>= \! \! {\small \prod_{s=1}^N} \! \!
<X_s;t|X'_s;0> \! \! .
$$ 
The transformation  being linear and with Jacobian $1$, 
the Green's function in the original $x_i$ is simply 
equal to the new one, with $X_s=X_s(x_i)$.

 As the  integrations in Eq.(1) have to be done numerically 
we have been able to solve for up to $N=4$ particles. 
 The probability distribution is again some Gaussian-like curve that gets
broader with time (the zero mode corresponding to a collective
movement of the circle of strings is allowed and so the center
of mass of the system disperses as usual in QM). As we increase the tempera-
ture the coordinate $x_1$ disperses much faster due to the momentum transfer
from the other oscillators. It would be interesting in the future to study
to which extent we have reached the thermodynamic limit in the sense that
if we have a very large circle of $N$ oscillators the probability to find the
first in some position after some time, should converge to some limit which 
should not change when we go to $N+1$ oscillators.\\
\vspace{-2mm}
\section{Conclusions}
  We have been able to combine non perturbatively real time and finite
Temperature in first quantization. 
 It can be seen as the product of three Green functions in the complex
time plane. The "matrix iteration" and the FFT methods 
work well in calculating the real time path integral, improving 
convergence with a small $i \epsilon$.   
  
 There are several interesting problems that seem shortly doable, like
the ammonia molecule or several spin models (easy due to discrete field values).
 Our ultimate aim is to find a way to treat the field theory case of tunneling
with instantons in real time and finite $T$, which is in principle
doable with these methods but will be hard to implement due
to the number of degrees of freedom.

 We thank L. Polley for useful discussions and the RRZN in Hannover for
supercomputer time.
 
\end{document}